\documentclass[12pt,preprint]{aastex}

\begin{document}

\title{Detecting the Attenuation of Blazar Gamma-ray Emission
 by Extragalactic Background Light with GLAST}

\author{Andrew Chen}
\email{chen@mi.iasf.cnr.it}
\affil{Consorzio Interuniversitario per la Fiscia Spaziale, \\
Torino, Italy  10133}
\and
\affil{Istituto di Astrofisica Spaziale e Fisica Cosmica, Sezione di Milano,\\
Milano, Italy  20133}

\author{Luis C. Reyes}
\affil{University of Maryland at College Park,\\ College Park, MD  20742} 
\and
\affil{Laboratory for High Energy Astrophysics, NASA/Goddard Space
Flight Center,\\ Greenbelt, MD  20771}
\author{Steven Ritz}
\affil{Laboratory for High Energy Astrophysics, NASA/Goddard Space
Flight Center,\\ Greenbelt, MD  20771}

\begin{abstract}

Gamma rays with energy above 10~GeV interact with optical-UV photons
resulting in pair production. Therefore, a large sample of high redshift
sources of these gamma rays can be used to probe the extragalactic
background starlight (EBL) by examining the redshift dependence of
the attenuation of the flux above 10~GeV. GLAST, the next generation
high-energy gamma-ray telescope, will have the unique capability to
detect thousands of gamma-ray blazars to redshifts of at least $z=4$,
with sufficient angular resolution to allow identification of a large
fraction of their optical counterparts. By combining established models
of the gamma-ray blazar luminosity function, two different calculations
of the high energy gamma-ray opacity due to EBL absorption, and the
expected GLAST instrument performance to produce simulated fluxes
and redshifts for the blazars that GLAST would detect, we demonstrate 
that these gamma-ray blazars have the potential to be a highly effective 
probe of the optical-UV EBL.
\end{abstract}


\section{Introduction \label{sec-1}}

In the last few years the study of galaxy formation and evolution
has seen tremendous progress. Instruments at many different wavelengths
have begun to penetrate to the relevant redshifts. One important prediction
of models of galaxy formation and evolution is the nature of the radiation
field produced by star formation. One way to probe the resulting extragalactic
background light (EBL) is to measure the attenuation through pair
production of gamma rays from distant sources. However, without a
large sample of sources distributed across a wide redshift range,
it is difficult to distinguish between extragalactic absorption and
characteristics peculiar to individual sources. The Large Area 
Telescope (LAT) instrument on The Gamma-Ray Large Area Space Telescope 
(GLAST) will observe gamma rays with energies
from 20~MeV to $>300$~GeV. The GLAST LAT will be the first instrument able
to probe the intergalactic radiation field by observing the absorption
of gamma rays from a large number of extragalactic point sources as
a function of redshift over a wide range. Ground-based telescopes
can measure the attenuation of TeV emission by intergalactic IR radiation
(Stecker, DeJager, \& Salamon 1992; Macminn \& Primack 1996; Madau
\& Phinney 1996). However, these telescopes will measure the spectra of
relatively small number of sources, making it more difficult to resolve
the question of whether differences between sources are due to
intergalactic attenuation or intrinsic peculiarities. Furthermore,
the high pair production opacity of the IR radiation limits TeV
probes of the EBL to a narrow, low-redshift range. GLAST, on the other
hand, will observe thousands of sources, and will measure less drastic
attenuation of GeV photons by optical and UV radiation. The energy range
and capabilities of GLAST are thus ideal for probing the EBL to
cosmological distances.

This paper reports our first modeling of the ability of GLAST to measure
the extragalactic background light absorption. In order to do this,
we need 1) models of the intergalactic radiation field, 2) the luminosity
function of extragalactic gamma-ray sources, and 3) parameters of
the instrument. In Section 2.1 we briefly review models for the intergalactic
radiation field and the resulting gamma-ray opacity as a function
of redshift. In Section 2.2, we describe the two gamma-ray blazar
luminosity functions used. In Section 2.3 we describe the parameters
used to simulate GLAST. In Section 3 we discuss the simulation procedure,
including the two different models of blazar input spectra and the two
models for the intergalactic radiation field. In Section 4 we present our
results and conclusions.

\section{Framework \label{sec-2}}

\subsection{Extragalactic Background Light \label{sec-2.1}}

Gamma rays with $E>10$~GeV traveling through intergalactic space
will interact through pair production with the extragalactic background
starlight (EBL) emitted by galaxies. The total center-of-mass energy
must be high enough to produce the electron-positron pair, and, for
a wide range of EBL models, the attenuation becomes significant only
above $\sim10$~GeV. The cross section is maximized when the EBL photon
energy $\epsilon_{EBL}\sim\frac{1}{2}(\frac{1000\rm~GeV}{E_{\gamma}})$
eV, with $E_{\gamma}$ in GeV (Stecker, DeJager \& Salamon 1992).
For 10~GeV to TeV gamma rays this corresponds to $\epsilon_{EBL}$
in the optical-UV range. Salamon \& Stecker (1998) calculated the
opacity of high energy gamma-rays to redshift $z=3$. To estimate
the stellar emissivity and spectral energy distributions vs. redshift
they adapted the analysis of Fall, Charlot, \& Pei (1996), consistent
with the Canada-France Redshift Survey, and included corrections for
metallicity evolution. They found that the stellar emissivity peaks
between $z=1$ and 2 before falling off, leading to a significant 
redshift-dependent absorption below $z=3$. Other models, \textit{e.g.} by 
Primack et al. (1999), provide for significant attenuation at even larger 
redshifts.  More recently, Bernstein, Freedman, \& Madore (2002a,b)
have made the first direct measurement of the optical-UV EBL integrated 
over redshift.  As shown in Section 3, our technique is a powerful discriminator among
models, giving information about the era of galaxy formation and evolution.

\subsection{Gamma-ray Blazars \label{sec-2.2}}

\subsubsection{Blazar Luminosity Function \label{sec-2.2.1}}

The Energetic Gamma Ray Experiment Telescope (EGRET) detected more
than 60 blazar-type quasars (Mukherjee et al. 1997) emitting gamma
rays with $E>100$~MeV. These sources are flat-spectrum radio-loud
quasars (FSRQs) and BL Lac objects, often exhibiting non-thermal radio
continuum spectra, violent optical variability and/or high optical
polarization. They are also highly variable and powerful gamma-ray
sources. The EGRET blazars whose optical redshifts have been measured
lie between $z=0.03$ and 2.28. The redshift distribution is consistent
with the observed distribution of FSRQs, which extends up to $z=3.8$.
However, since the luminosity function determines the statistical power of
our technique versus redshift, and since this function is still relatively
unconstrained, we use two different models for the blazar luminosity
function.

The first model, by Salamon \& Stecker (1996), makes the assumption that
blazars seen in gamma-rays above 100 MeV are also seen in the radio
as FSRQs. This model assumes that the gamma and radio ray
luminosity functions are linearly related as\[
\rho_{\gamma}\left(L_{\gamma},z\right)=\eta\rho_{r}\left(L_{r},z\right)\]
 where $\eta$ is a parameter of the model and \[
\rho_{r}\left(L_{r},z\right)=10^{-8.15}\left\{ \left(\frac{L_{r}}{L_{c}\left(z\right)}\right)^{0.83}+\left(\frac{L_{r}}{L_{c}\left(z\right)}\right)^{1.96}\right\} ^{-1}\]
with $\log_{10}L_{c}\left(z\right)=25.26+1.18z-0.28z^{2}$. The units
of the comoving density $\rho$ are Mpc$^{-3}\times\left(\right.$unit interval
of $\log_{10}L\left.\right)^{-1}$ and the units of $L$ are
W Hz$^{-1}$ sr$^{-1}$.  Using cosmological parameters $\Omega_M=1, \Omega_{\Lambda}=0,$
and $H_0 = 50$ km/s/Mpc, the model is constrained to predict the number of 
blazars observed by EGRET.

The number of sources with redshift in the interval $z+\Delta z$ seen at
Earth with a flux for $E > 100$ MeV in the interval $F+\Delta F$ is 
given by (Salamon \&
Stecker 1996)\[\frac{dN}{dFdz}\Delta z\Delta F=4\pi R_{0}^{3}r^{2}\Delta r\rho_{\gamma}\Delta\left(\log_{10}L\right)\]
with $R_{0}r=\frac{2c}{H_{0}}\left(1-\left(1+z\right)^{-1/2}\right)$,
where $H_{0}$ is the Hubble expansion rate. Combining the choice of
parameters given by Salamon \& Stecker (1996)
with a GLAST flux sensitivity of $1.5\times10^{-9}$
photons cm$^{-2}$s$^{-1}$, a number of $\sim9000$
blazars is expected to be observed with redshifts up to $z\sim4$.

The second model, by Chiang \& Mukherjee (1998), does not assume a
correlation between luminosities at gamma ray energies and at other
wavelengths. This model parametrizes the luminosity function as
\begin{eqnarray*} \frac{dN}{dL_{0}dV}\propto\left(\frac{L_{0}}{L_{B}}\right)^{-\gamma_{1}} & , & L_{0}\leq L_{B}\\
\frac{dN}{dL_{0}dV}\propto\left(\frac{L_{0}}{L_{B}}\right)^{-2.2} & , & L_{0}>L_{B}\end{eqnarray*}
with de-evolved luminosity $L_{0}=L/\left(1+z\right)^{\beta}$ and
a maximum cutoff redshift of $z_{max}=5$. The energy range of this
integrated luminosity is $E > 100$ MeV. The best fit found for this
broken power law is parametrized by $\gamma_{1}\lesssim1.2$,
$L_{B}=1.1\times10^{46}$ erg/s
and $\beta=2.7$, with cosmological parameters $\Omega_M=1, \Omega_{\Lambda}=0,$
and $H_0 = 75$ km/s/Mpc.  Each model was separately fit in a self-consistent
fashion to the EGRET data to produce the luminosity functions.  More recent
cosmological data suggest a non-zero value for $\Omega_{\Lambda}$. The impact 
on the luminosity function, however, is small; we therefore retain the original
model, along with the 
fit to the data, for our calculations.  Of course, one of the important goals
of the GLAST mission will be to constrain the luminosity function.

\subsubsection{Blazar Spectra \label{sec-2.2.2}}

The spectra of the blazars observed by EGRET are well characterized
in the $E > 100$~MeV range by power laws with an average photon spectral
index of $-2.15\pm0.04$ (Mukherjee et al. 1997). The spectra of some
individual blazars have a measured index significantly different from
the mean value, suggesting true scatter in the distribution of blazar
spectra, which our simulation takes into account as described below.
More importantly, most of the EGRET blazars have not been detected
by TeV telescopes; for many of these sources, this implies a spectral
break or rolloff at some energy between the EGRET and TeV energy ranges.
Intergalactic attenuation, the very effect explored in this paper,
would account for the lack of detection of high-redshift objects,
but there are relatively low-redshift blazars that are bright in the
EGRET range and undetected in the TeV range. More tellingly, most of
the TeV blazars belong to the same subset of blazars, the X-ray selected
BL Lac objects (XBLs). Since only a small fraction of the EGRET blazars
are XBLs, this implies that the non-XBL blazars may have spectra with
intrinsic rolloffs independent of any intergalactic attenuation effects.
Finally, blazars that have been detected in both the GeV and TeV ranges
have TeV fluxes that are lower than simple extrapolations of the EGRET
power laws would suggest. Of course, such an extrapolation over such
a wide range of energies is unreasonable. Most of the models for blazar
spectra attribute both the GeV and TeV emission to the same inverse
Compton component of the emission. However, with little observational
data in the 30-300 GeV range, no firm conclusions can be drawn about
the precise shape of the spectra. Indeed, this is one of the motivations
for the next generation of experiments.

Our technique, as described in Section \ref{sec-3}, is to form the ratio
of the observed fluxes for $E\,>\,10\,$GeV and $E\,>\,1\,$GeV,
\[\frac{F\left(E\,>\,10\,\rm~GeV\right)}{F\left(E\,>\,1\,\rm~GeV\right)}\] This
ratio is simple, robust, and insensitive to rolloffs above $\sim50$~GeV
for most EBL models as shown in Section 3.1. We attempt to bracket the
range of possible spectra by first analyzing a sample of blazars whose
power law spectral indices are normally distributed around a mean of -2.15
with standard deviation 0.04, representing a situation where there is a
range of spectral indices but no intrinsic rolloff in this energy range.
To model intrinsic rolloffs, we then repeat the analysis with a sample of
blazars whose unredshifted spectra have a broken power law with mean index
-2.15 below  50~GeV and -3.15 above, again with a standard deviation of
0.04 in each case.

\subsection{GLAST \label{sec-2.3}}

The Gamma-ray Large Area Space Telescope (GLAST) is under development
with a planned launch in 2006 (Michelson 2001). The
Large Area Telescope (LAT) of GLAST
will observe gamma rays with energies from 20~MeV to $>300$~GeV.
GLAST will have a much larger effective area than EGRET, especially
at higher energies (peak effective area $>8000$ cm$^{2}$ at $>1$~GeV), 
a larger field of view, and sub-arcminute scale source localization.
GLAST should be able to reach a $5\sigma$ point source flux sensitivity
of less than $1.5\times10^{-9}$ photons cm$^{-2}$ s$^{-1}$ for
$E>100$~MeV within five years. As noted above, using the distribution of
blazars observed by EGRET and extrapolating to lower fluxes,
it is estimated that GLAST will detect thousands of blazars. Improved
angular resolution should allow a high percentage of optical identifications
and redshift measurements, depending on the available ground-based
resources. Improved high-energy performance should yield accurate
flux determinations above 10~GeV for many of these sources. Note
that our modeling is based on the generic parameters outlined in the
GLAST Science Requirements Document (Michelson 2001);
the performance of the flight instrument may be substantially better.

\section{Procedure \label{sec-3}}

To simulate the gamma-ray sources observable by GLAST, we need a reasonable
extrapolation of the EGRET source distribution to the GLAST flux limit. We
used the two luminosity functions described in section \ref{sec-2.2.1}
for this purpose, but our main conclusions do not depend significantly
on this choice. We note that any predictions made now will be supplanted
by the data GLAST itself provides.

Before any observational selection, according to the luminosity function
by Salamon \& Stecker (1996), $\sim12,000$
blazars in principle will have fluxes in the range detectable by GLAST.
Each one was assigned a random luminosity and redshift according to
this model.

With the luminosity function by Chiang \& Mukherjee (1998) we generated
10,000 blazars, between redshifts
0 and 5 according to Figure 6 of their paper.

For both samples, the flux of each blazar was then calculated according to
\[F=L\frac{(1+z)^{2-\alpha}}{4\pi d_{l}^{2}(z)},\]
where $\alpha$ is the photon spectral index and $d_{l}$ is the cosmological
luminosity distance $d_{l}=\frac{2c}{H_{0}}\left(1+z\right)\left[1-\left(1+z\right)^{-1/2}\right]$.

Only blazars with observed flux greater than $1.5\times10^{-9}$
photons cm$^{-2}\, s^{-1}$
for $E>100\, MeV$ are allowed in the sample. The $E>10$~GeV flux of
each blazar was calculated by adding two effects. First, each blazar
was given a random, normally distributed spectral index, $-2.15\pm0.04$.
An index of -2.15 yields a flux ratio
$\frac{F\left(E\,>\,10\,\rm~GeV\right)}{F\left(E\,>\,1\,\rm~GeV\right)}$
of $\sim0.07$. Also included was the
redshift-dependent absorption above 10~GeV.
The form of the dependence was parameterized from Figure 6 of Salamon
\& Stecker (1998), with metallicity corrections.  In this EBL model, $\Omega_M=1$, 
$\Omega_{\Lambda}=0$, and the value of $H_0$ scales out.  We set the absorption
for $z>3$ for this model equal to the absorption at $z=3$, both
because it is a conservative assumption and because it is physically
plausible (little stellar emissivity, and smaller scale and path lengths,
for $z>3$). To produce observed fluxes from these intrinsic fluxes,
each blazar was assigned a random position on the sky and, assuming
an exposure equivalent to two full years, Galactic and extragalactic
backgrounds were added. The Galactic backgrounds were derived from
the diffuse model used in EGRET analysis (Hunter et al. 1997). To
take into account the extragalactic background, we added a second,
fixed, background component, with intensity $4\times10^{-6}$ photons
cm$^{-2}$ s$^{-1}$ sr$^{-1}$ for $E>100$~MeV and a power law
index of -2.15, under the assumption that GLAST may resolve a significant
fraction of the EGRET isotropic background (Stecker \& Salamon 1999).
This component represents the sum of the flux from unresolved blazars
and any truly diffuse background contribution. Any blazar within $10\arcdeg$
of the Galactic plane and any blazar whose observed flux was less
than $5\sigma$ above the background flux at $E>1$~GeV was removed
from the sample, leaving $\sim9100$ blazars (Stecker \&
Salamon 1996) or $\sim8200$ blazars (Chiang \& Mukherjee
1998). Figure 1 shows a histogram of the number of blazars in each
0.5 redshift bin. The model by Chiang \& Mukherjee (1998) predicts
a population of blazars that are intrinsically brighter when compared
to the model by Stecker \& Salamon (1996). In that case, GLAST would
detect more blazars at higher redshift as can be observed from the
graph. We note that, with no EBL attenuation, for $z>3$ and requiring more 
than 5 detected photons $(E\,>\,10\,\rm~GeV)$, GLAST would see $\sim60$ blazars using the 
Stecker \& Salamon luminosity function, or $\sim700$ blazars using the Chiang \&
Mukherjee luminosity function.

\placefigure{fig1}

\subsection{Calculating the flux ratios \label{sec-3.2}}

The integrated fluxes of each blazar for $E>1$~GeV and $E>10$~GeV
were used to generate observed fluxes using Poisson distributions
equivalent to two full years of exposure. For each blazar, we calculated
the ratio between these fluxes. The error in each flux ratio was set
to
$\sigma_{ratio}=\frac{1}{F(E\,>\,1\,\rm~GeV)}\sqrt{{\sigma_{F(E\,>\,10\,\rm~GeV)}}^{2}+(\frac{F(E\,>\,10\,\rm~GeV)}{F(E\,>\,1\,\rm~GeV)}\sigma_{F(E\,>\,1\,\rm~GeV)})^{2}}$,
where $\sigma_{F}$ is the statistical error of the flux measurement
in each energy range.  The crosses in Figure~\ref{fig2} show
the weighted mean ratio in each redshift bin. To avoid the bias of small
number Poisson statistics toward lower values, the flux ratio of each
source was weighted by the Poisson error of the $E>1$~GeV flux,
rather than the formal, propagated error of the flux ratio. The
diamonds show the same ratio when the intergalactic absorption is
removed from the observed blazar fluxes. In all cases the error bars
are statistical, obtained by computing the rms scatter within each redshift
bin and dividing by $\sqrt{N}$. The analytically derived flux ratio using
the opacity
model of Salamon \& Stecker is plotted as a solid curve. For comparison,
the dashed lines in Figure~\ref{fig2} show the same results with no
intergalactic absorption.

We repeated the entire analysis with the blazar spectra changed from
single power laws with mean index -2.15 to broken power laws with
mean index -2.15 below 50~GeV (at the source) and -3.15 above. The results
are plotted as crosses in Figure~\ref{fig3}. Although fewer blazars have
detected flux above 10~GeV, the effects of absorption are still apparent.
Note that sources with no detectable flux above 10~GeV (zero photons)
still provide important information; indeed, neglecting them introduces
a bias. The modified $\chi^{2}$ statistic used here (Mighell 1999)
accounts for these sources.

The ratio obtained without EBL absorption is presented as diamonds, along with
the analytically derived flux ratio (dashed line). As can be easily seen, this
flux ratio is not constant as a function of redshift. This is a consequence of
defining the break in the index for a given energy at the source.

\placefigure{fig2}
\placefigure{fig3}

\subsection{Other EBL models \label{sec-3.3}}

Primack and collaborators combined theoretical modeling
with observational data to develop semi-analytic models
of galaxy formation and evolution (Primack et al. 1999). Their models
permit a physical treatment of the processes of galaxy formation and
evolution in a cosmological framework, including gravitational collapse,
mergers, etc., rather than relying on pure luminosity evolution of
the galaxies existing today.  We use their calculations 
of opacities to gamma rays at redshifts up to $z=5$. The cosmological parameters used 
are $\Omega_M=0.4$, $\Omega_{\Lambda}=0.6$, and $H_0=60$ km/s/Mpc. 
The luminosity functions use a different value for $H_0$, but for our 
purposes this difference does not significantly affect the results; as 
shown by Blanch and Martinez (2003), the gamma-ray horizon has a 
relatively weak dependence on $H_0$. Note that these opacities, with their 
different cosmological parameter sets, should not be thought of as 
predictions, but rather as another set of reasonable values to illustrate 
the discriminating power of our technique. The results are shown as triangles in Figure~\ref{fig2}
and Figure~\ref{fig3}, along with the lines representing the analytical
prediction. The fact that the flux seems to be more highly attenuated
is not important. What is more interesting is that the decrease in
flux ratio from $z=2.5$ to $z=5$ is observable. This indicates, assuming
the availability of gamma-ray sources and sufficient EBL density,
that EBL absorption can effectively probe galaxy formation at those
redshifts, a regime of intense theoretical interest.
More recently, Oh (2001) performed an independent calculation of the opacity of
gamma-ray blazar emission to pair production by UV photons as a function of 
redshift.  While not addressing the detectability of high-redshift blazars by 
GLAST in detail, he obtains attenuation factors that vary strongly with
redshift in a manner roughly consistent with the calculations we have used.

\section{Results and Conclusions \label{sec-4}}

Extragalactic attenuation of gamma-rays by low-energy background photons
produces a distortion in the spectra of gamma-ray blazars that increases
with increasing redshift. Because we cannot distinguish the difference
between extragalactic attenuation and intrinsic effects in individual
blazar spectra, statistical analysis of a large sample of blazars
such as those presented in this paper is a powerful tool to study
EBL absorption. Although AGILE, the next GeV mission (Tavani et al. 2001), 
will produce a significant
increase in the total number of blazars and therefore refine the blazar
luminosity function and evolution, GLAST will be the first mission to
observe a large sample of high redshift blazars with sufficient statistics 
to separate intrinsic differences between blazars from redshift dependence 
of EBL absorption.  Our results indicate that the
redshift dependence of the attenuation should be easily detectable
by GLAST even when the diffuse background is taken into account and
possible high-energy intrinsic rolloffs are considered.

Selection effects, both from GLAST itself and from optical coverage
of redshift determinations, will primarily affect sources with low
flux. These sources will have poorly measured flux ratios, and will
suffer from optical selection effects due to their more poorly determined
positions. Other biases include the locations of optical telescopes,
source clustering, and other effects. It will be important to catalog
these effects explicitly; in particular, insuring adequate optical
coverage may require active preparation and participation.

GLAST will be able to measure the differences in blazar attenuation
in the cosmologically interesting range in redshift from $z=1$ up
to $z=5$. This is in contrast to ground-based observations of TeV
attenuation by IR radiation, which will only be able to measure 
differences well below $z=1$, where the IR becomes opaque.
As the energy threshold of the ground based experiments
drops over time, their redshift range will increase, but will remain 
limited to low redshifts except for exceptional, statistically 
insignificant special cases, especially given their generally small 
fields of view.
More than establishing that EBL attenuation occurs, GLAST will be
able to distinguish between different EBL models. This would validate
EBL attenuation as a direct cosmological probe.

We emphasize that this analysis will require redshift determinations
of a large fraction of GLAST blazars. This is another example of the
importance of cross-wavelength studies: by using optical measurements
of blazar redshifts, gamma-ray measurements can uniquely probe the
optical-UV EBL. A redshift measurement for thousands of high-redshift
sources is not a trivial undertaking, but the effort will be well
rewarded.

Even after observation of a redshift-dependent effect, the possibility
would remain that the spectral evolution of gamma-ray blazars might
coincidentally mimic redshift-dependent EBL absorption. For example,
if blazars that formed in the early universe suffered more internal
attenuation than blazars that formed later, the same effect could
be produced. Note that blazars are variable, and there are some indications
that their spectra can become harder when they flare (Sreekumar et
al. 1996). Evolution in flaring probability could produce the same
effect as actual spectral evolution from a statistical standpoint
(for example, a higher percentage of high-redshift blazars might be
observed in a quiescent phase), although one would expect the GLAST
flux limit to produce a selection effect in the opposite direction.
In any case, observation of a redshift-dependent spectral softening
will provide an important constraint. Theorists will have to decide
the likelihood of an evolutionary conspiracy.

\acknowledgments

We acknowledge useful conversations with Bill Atwood, who first suggested using
the large statistics of GLAST AGNs to look for systematic effects of
extragalactic background light attenuation with redshift.  We would also like to
thank David Thompson, Seth Digel, Floyd Stecker, and Mike Salamon
for their useful comments, and James Bullock and Joel Primack for providing
calculations from their models.  We would like to thank the referees for
many useful comments, in particular drawing our attention to the
Blanch \& Martinez article on cosmological effects in emissivity
evolution.

\clearpage

\figcaption[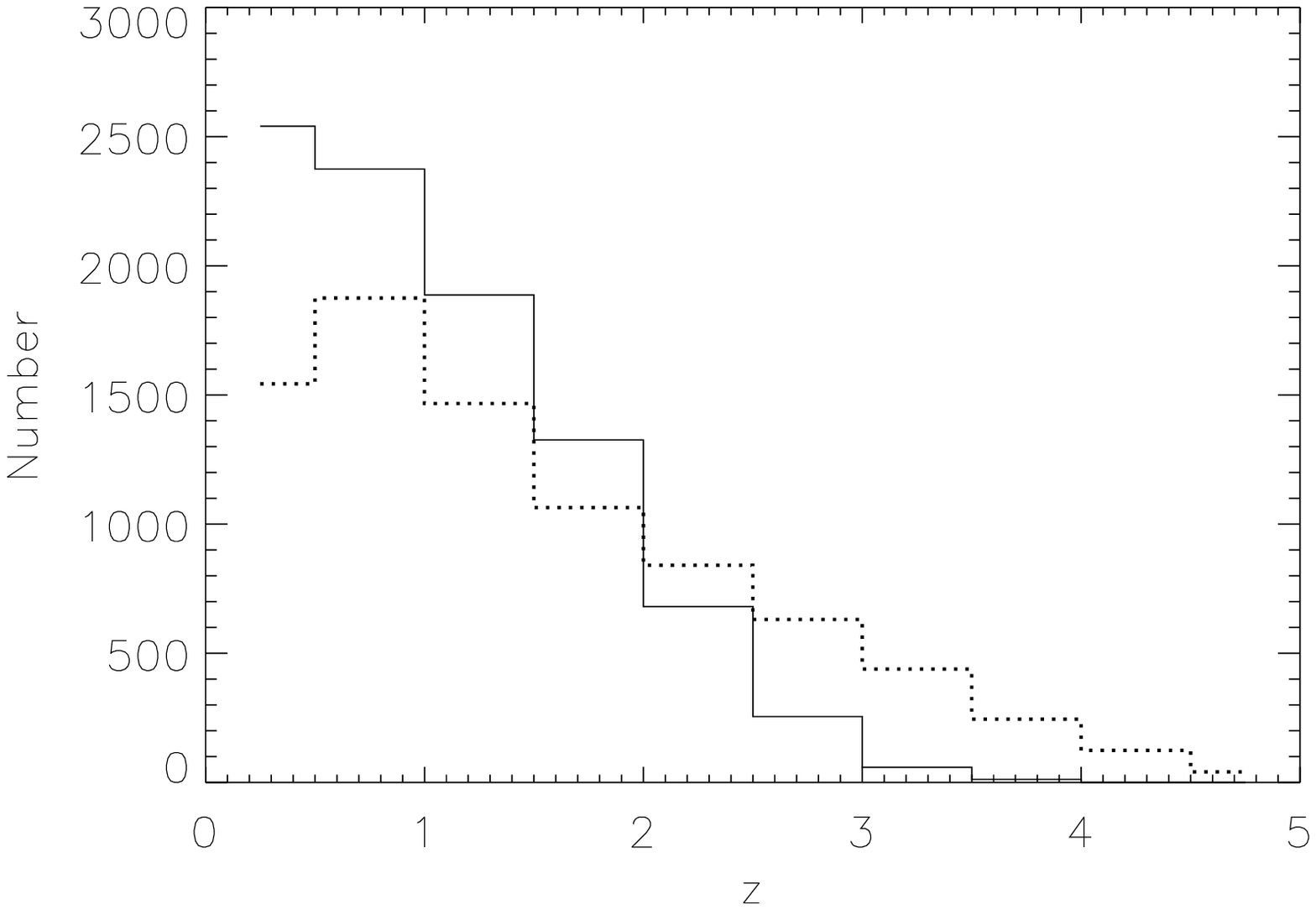]{Number of detectable blazars in each redshift
interval. The solid curve is the population according to Stecker \& Salamon (1996),
the dotted line corresponds to the model by Chiang \& Mukherjee (1998).
The spectra of the blazars are power laws with index
\(-2.15 \pm 0.04\).
\label{fig1}}

\figcaption[f2.ps]{Mean observed flux ratio, as described in the
text, using the luminosity function by
Salomon \& Stecker (a) and Chiang \& Mukherjee (b). Each cross is the mean observed
flux ratio in the corresponding redshift interval with fluxes attenuated by the EBL
of Salamon \& Stecker.  The solid curve is the ratio calculated with the same model.
The triangles and the dash-dot-dotted line are the mean observed
and calculated flux ratios for blazars, with the EBL attenuation model of Primack
et al. Finally, the diamonds show the mean observed flux ratio with
no EBL attenuation and the dashed line is the corresponding calculated ratio.
\label{fig2}}

\figcaption[f3.ps]{Mean observed flux ratio as described in the text
for blazars with broken spectral index at 50~GeV at the source. The
luminosity function is obtained from  Stecker \& Salamon (a) and
Chiang \& Mukherjee (b). The EBL attenuation is given by either Salamon \& Stecker
(crosses) or Primack et al (triangles). The analytically calculated flux ratios for each
luminosity function are shown by the solid and dash-dot-dotted lines respectively. The dashed
line and the diamonds show the same results when there is no EBL attenuation.\label{fig3}}

\clearpage

\plotone{f1.eps}

\epsscale{0.8}
\plotone{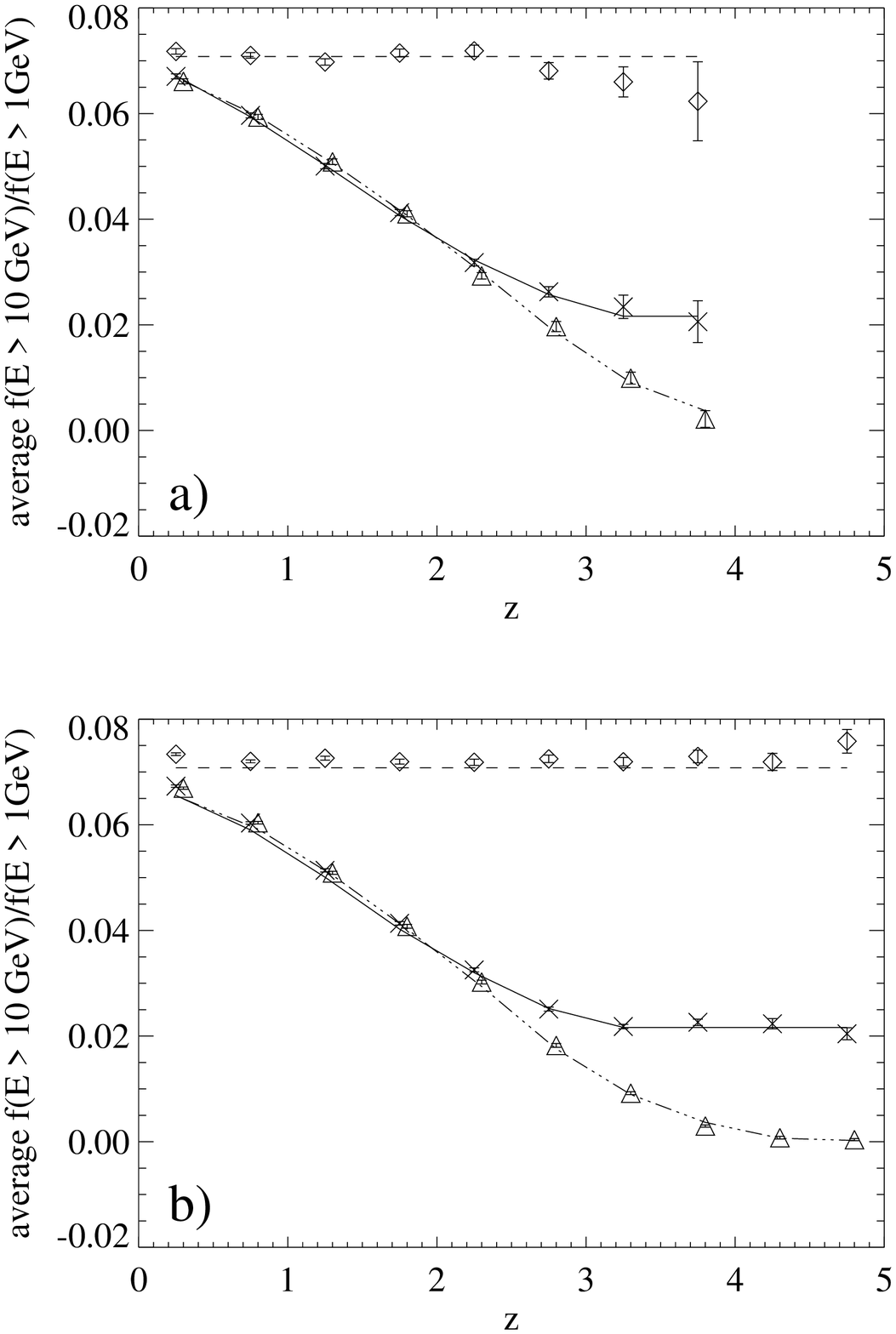}

\epsscale{0.8}
\plotone{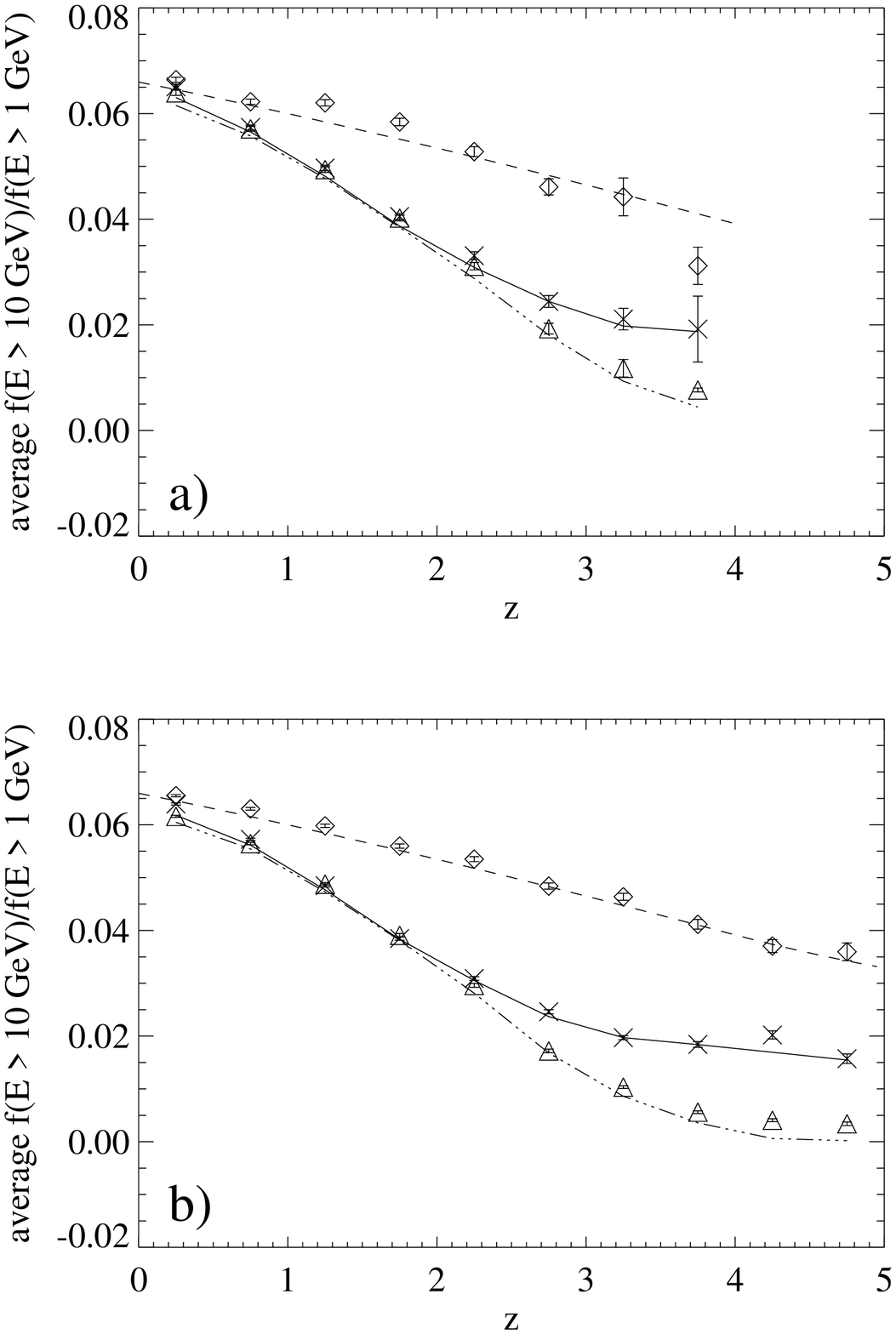}

\end{document}